\documentclass[pra,twocolumn,amsmath,amssymb]{revtex4-1}

\usepackage{epsfig,amsmath}
\usepackage{subfigure}
\usepackage{graphicx}
\usepackage{dcolumn}
\usepackage{stmaryrd}
\usepackage{mathrsfs}
\usepackage{pifont}
\usepackage{amsthm}
\usepackage{amssymb}
\usepackage{bm}
\usepackage{latexsym}
\usepackage[colorlinks=true,linkcolor=blue,citecolor=blue]{hyperref}
\usepackage{color}
\usepackage{epstopdf}

\theoremstyle{plain}

\newcommand{\ti}{\tilde}

\newcommand{\De}{\Delta}

\newcommand{\om}{\omega}

\newcommand{\non}{\nonumber}

\begin{document}

\title{Criterion for quantum Zeno and anti-Zeno effects}

\author{Jia-Ming Zhang}
\author{Jun Jing}\email{Email address: jingjun@zju.edu.cn}
\author{Li-Gang Wang}
\author{Shi-Yao Zhu}

\affiliation{Department of Physics, Zhejiang University, Hangzhou 310027, Zhejiang, China}

\date{\today}

\begin{abstract}
In this work, we study the decay behavior of a two-level system under the competing influence of a dissipative environment and repetitive measurements. The sign of the second derivative of the environmental spectral density function with respect to the system transition frequency is found to be a sufficient condition to distinguish between the quantum Zeno (negative) and the anti-Zeno (positive) effects raised by the measurements. We check our criterion for practical measurement intervals, which are larger than the conceptual Zeno time, in various environments. In particular, with the Lorentzian spectrum, the quantum Zeno and anti-Zeno phenomena are found to emerge respectively in the near-resonant and off-resonant cases. For the interacting spectra of hydrogenlike atoms, the quantum Zeno effect usually occurs and the anti-Zeno effect can rarely occur unless the transition frequency is close to the cut-off frequency. With a power-law spectrum, we find that sub-Ohmic and super-Ohmic environments lead to the quantum Zeno and anti-Zeno effects, respectively.
\end{abstract}

\maketitle

\section{Introduction}\label{intro}

No quantum systems can be completely isolated from their environment. Every quantum system should be regarded as an open system due to the unavoidable coupling with the surrounding environment. Mutual interaction between system and environment is responsible for many important physical processes, such as dissipation, fluctuation, and pure decoherence~\cite{Breuer2002,Nielsen2000}. The theory of open-quantum systems has attracted growing attentions in the fields of quantum noise, quantum optics, and quantum information science~\cite{Gardiner2004,Scully2012}.

The decay of any unstable quantum state can be inhibited by repeated measurements in a very frequent way. This ``watchdog'' phenomenon, known as the quantum Zeno effect (QZE), is named after the pre-Socratic Greek philosopher Zeno, who argued that an arrow in flight, if observed, does not move. The QZE has been proposed to suitably confine the evolution of quantum systems and attracts theoretical and experimental interest. It has been used to protect quantum information~\cite{Barenco1997}, to preserve quantum coherence~\cite{Beige2000,Search2000,Zhou2009} and entanglement~\cite{Maniscalco2008,Wang2008}, to cool down and purify a quantum system~\cite{Erez2008}, to suppress intramolecular forces~\cite{Wuster2017}, and to realize direct counterfactual communication~\cite{Cao2017}. In contrast, the measurements can also speed up the temporal evolution of quantum systems, when the measurements are not sufficiently frequent. This Heraclitus (who replied that everything flows) behavior, i.e., the opposite of the QZE, is called the quantum anti-Zeno effect (QAZE). Moreover, the QAZE is predicted to be an even more ubiquitous phenomenon~\cite{Kaulakys1997,Kofman2000}. Both the QZE and the QAZE have been experimentally observed in many physical systems, such as trapped ions~\cite{Itano1990} and trapped atoms~\cite{Fischer2001}, superconducting qubits~\cite{Barone2004,Harrington2017,Kakuyanagi2015,Slichter2016}, Bose-Einstein condensates~\cite{Streed2006}, nanomechanical oscillators~\cite{Chen2010}, cavity quantum electrodynamics systems~\cite{Helmer2009}, and nuclear spin systems~\cite{Wolters2013,Zheng2013,Kalb2016}.

The QZE and QAZE dynamics have been studied in various open-quantum-system scenarios, where the system is subject to population decay~\cite{Segal2007,Chaudhry2016,He2017,Zhou2017}, pure dephasing~\cite{Chaudhry2014,Zhang2015}, and quantum Brownian motion~\cite{Maniscalco2006}, respectively. A conventional test-bed is a two-level atom coupled to a vacuum-field reservoir. By applying a conventional unitary transformation (UT) approach, Kofman and Kurizki found that the modification of the decay process is determined by the spectral density function (SDF) of the environment and the energy spread induced by the measurements~\cite{Kofman1996,Kofman2000}. This method has also been applied to investigate the QZE and QAZE in different structured environments~\cite{Zheng2008,Wang2009,Li2009,Cao2010,Ai2010}.

In general, by changing the time spacing between successive measurements, the decay can be suppressed or accelerated, depending on the features of the interaction Hamiltonian. According to Refs.~\cite{Facchi2001,Koshino2005,Facchi2008}, to obtain a criterion for the Q(A)ZE, considerable efforts have to be made to scrutinize the general features of the system-environment interaction and calculate the residue of the propagator involving the environmental SDF. To the best of our knowledge, the conditions for probing the QZE or the QAZE have not been explicitly and simply stated. The target of this work is to propose a concisely analytical criterion to predict the QZE or the QAZE. It is found that the sign of the second derivative of the environmental SDF with respect to the system transition frequency is sufficient to discriminate between the quantum Zeno (negative) and the anti-Zeno (positive) effects qualitatively.

The rest of this work is organized as follows. In Sec.~\ref{gener}, we shall briefly introduce the effective decay rate based on the UT approach. Section~\ref{concavity} is devoted to deduce our criterion or probe for the QZE or the QAZE. In Sec.~\ref{nonMar}, we employ this criterion to study the QZE or the QAZE in several structured environments. In Sec.~\ref{dissc}, we will discuss the range of applications of our criterion. We close this work with a summary in Sec.~\ref{concl}.

\section{The QZE and the QAZE in a general damped Jaynes-Cummings model}\label{gener}

We start by briefly recalling the theoretical derivation of the quantum Zeno and anti-Zeno effects in a two-level system (TLS) undergoing decay into a zero-temperature environment. The full Hamiltonian describing this decay process, with the rotating-wave approximation (RWA) and in units of $\hbar$, reads
\begin{equation}\label{H0}
H=\Delta\sigma^+\sigma^-+\sum_k\omega_k b_k^{\dag}b_k +\sum_k\left(g_k^*b_k\sigma^++g_kb_k^{\dag}\sigma^-\right),
\end{equation}
where $\Delta$ is the energy difference between the two levels, $\sigma^+$ and $\sigma^-$ are respectively the raising and lowering operators, $b_k^\dag$ and $b_k$ are respectively the creation and annihilation operators for the $k$-th mode of the environment with frequency $\omega_k$, and $g_k$ describes the coupling strength between the system and the $k$-th mode of the environment. If instantaneous projections are performed with a sufficiently small interval $\tau$, after $N$ periodical measurements, the survival probability becomes
\begin{equation}
p(t=N\tau)\equiv e^{-\gamma(\tau)t},
\end{equation}
where $t=N\tau$. Here the decay rate $\gamma(\tau)$ is dependent on $\tau$. Based on the UT approach, the measurement-modified decay rate is given by~\cite{Kofman2000}:
\begin{equation}
\label{gamma}
\gamma_{\rm eff}(\tau)=2\pi\int_0^\infty d\omega G(\omega)F(\omega,\tau),
\end{equation}
where $G(\omega)=\sum_k|g_k|^2\delta(\omega-\omega_k)$ is the environmental SDF and
\begin{equation}
\label{F}
F(\omega,\tau)=\frac{\tau}{2\pi}{\rm sinc}^2\left[\frac{(\omega-\Delta)\tau}{2}\right],
\end{equation}
is a ``filter function" describing the effect of the equidistant projections. The necessary details can be found in Appendix~\ref{UT}.

\begin{figure}[htbp]
\centering
\includegraphics[width=0.5\textwidth]{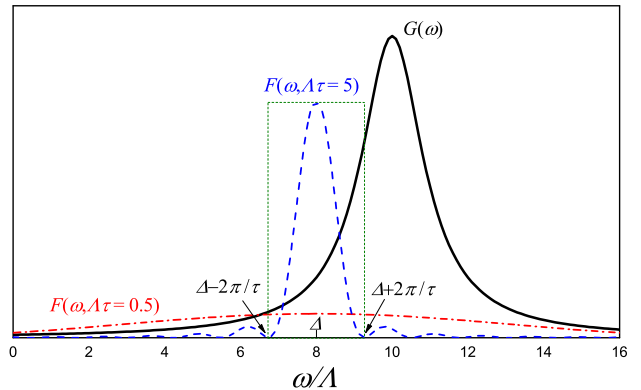}
\caption{(Color online) Schematic of the overlaps between the spectrum and different filter functions. The solid black line is the spectral density $G(\omega)$, which can be arbitrarily chosen. Here we plot a spectral density function of the Lorentzian form in Eq.~(\ref{LF}), centered at $\omega_0=10\Lambda$. The red dot-dashed line and the blue dashed line describe two filter functions $F(\omega,\tau)$ determined by the periodical measurements, which are centered at $\Delta=8\Lambda$ with a small measurement interval ($\Lambda\tau=0.5$) and a large one ($\Lambda\tau=5$), respectively. The green dotted frame is used to distinguish the main lobe of the filter function with $\Lambda\tau=5$. }
\label{schematic}
\end{figure}

As shown in Appendix~\ref{first}, we can also apply the first-order time-dependent perturbation theory to get the same result in Eq.~(\ref{gamma}). A general conclusion can be drawn from Eq.~(\ref{gamma}) that the effective decay rate modified by the repeated measurements is dependent on the overlap between the environmental SDF and a measurement-induced broadening of the TLS energy difference, as shown in Fig.~\ref{schematic}. The effective decay rate, which is insensitive to the total evolution time $t$, is determined by three factors: (i) the SDF $G(\omega)$, (ii) the transition frequency $\Delta$ of the TLS, and (iii) the measurement spacing $\tau$. For a larger $\tau$, the filter function ``picks'' up the near-resonant (on-shell) contribution of the interaction (see the blue dashed line in Fig.~\ref{schematic}). In contrast, for a smaller $\tau$, the filter function ``explores'' all possible intermediate states, where the collective response of the off-resonant modes should be taken into account~\cite{Debierre2015} (see the red dot-dashed line in Fig.~\ref{schematic}).

The effective decay rate $\gamma_{\rm eff}$ is the crucial physical quantity to identify the QZE and the QAZE in open-quantum-system dynamics. In particular, the QZE occurs if $\gamma_{\rm eff}(\tau)/\gamma_0<1$, and the QAZE does if $\gamma_{\rm eff}(\tau)/\gamma_0>1$~\cite{Koshino2005,Zheng2008}. Here $\gamma_0$ is the decay rate under the Weisskopf-Wigner approximation, characterizing a purely Markovian exponential decay [see Eq.~(\ref{gamma0})].

\section{Explicit criterion for the QZE and QAZE}\label{concavity}

Now we consider the behavior of $\gamma_{\rm eff}(\tau)$ with practical measurement intervals. The effective decay rate can be expressed as a combination of the measurement-independent part $\gamma_0$ and the measurement-dependent part $\gamma_1(\tau)$:
\begin{equation}\label{g0g1}
\gamma_{\rm eff}(\tau)=\gamma_0+\gamma_1(\tau),
\end{equation}
where
\begin{equation}\label{gamma0}
\gamma_0=2\pi G(\Delta)\int_0^{\infty}d\omega F(\omega,\tau)\approx2\pi G(\Delta),
\end{equation}
and
\begin{equation}\label{gamma1}
\gamma_1(\tau)=2\pi\int_0^{\infty}d\omega [G(\omega)-G(\Delta)]F(\omega,\tau).
\end{equation}
Actually, $\gamma_0$ is the natural decay rate which can be obtained by Fermi's golden rule~\cite{Cohen-Tannoudji1992}, i.e., $\gamma_0=2\pi G(\Delta)$. It is the spontaneous decay rate or free decay rate of the TLS. Consequently, the sign of the remainder $\gamma_1(\tau)$ given by Eq.~(\ref{gamma1}) will determine whether the QZE or the QAZE occurs in the system dynamics. The sign rather than the magnitude of $\gamma_1(\tau)$ is the main focus of our work.

When the system energy $\Delta$ is not extremely smaller than both the center of gravity of $G(\omega)$ and the cut-off frequency of the environment, the dominant contribution to the integral over $\omega$ in Eq.~(\ref{gamma}) is from the main lobe of $F(\omega,\tau)$, which ranges from $\Delta-2\pi/\tau$ to $\Delta+2\pi/\tau$ as shown in Fig.~\ref{schematic}. Notice that $2\pi/\Delta$ is a reliable and loose lower-bound rather than a definite infimum for $\tau$. Then one can approximately choose $[\Delta-2\pi/\tau,\Delta+2\pi/\tau]$ as the integration domain of Eq.~(\ref{gamma1}) and obtain the asymptotic behavior of the system up to the order of ${\it O}(\tau^{-2})$:
\begin{equation}\label{G2}
\gamma_1(\tau)\approx\tilde{\gamma}_1(\tau)=\frac{4\pi}{\tau^2}G''(\Delta)+{\it O}\left(\frac{1}{\tau^3}\right).
\end{equation}
Here $G''(\omega)$ is the second derivative of the SDF $G(\omega)$ with respect to frequency. The detailed derivation is provided in Appendix~\ref{cal}. We thus find that, as $\tau\to\infty$, which means no measurement is performed, $\gamma_{\rm eff}$ approaches $\gamma_0$ from below when $G''(\Delta)<0$, and from above when $G''(\Delta)>0$. Equation~(\ref{G2}) thus leads to an explicit and rather simple criterion. In particular, the second derivative of the SDF around the system transition-frequency can be used to distinguish between the quantum Zeno and the anti-Zeno effects in the modified dynamics of the system. The SDF with a negative second derivative $G''(\Delta)<0$, which is convex around $\Delta$, will give rise to the QZE. In contrast, a positive second derivative $G''(\Delta)>0$, which is concave around $\Delta$, will give rise to the QAZE. We find that the correction by Eq.~(\ref{G2}) on Fermi's golden rule~(\ref{gamma0}) captures the measurement effect in quality. With this criterion, one can judge the measurement would accelerate or suppress the decay of the system under decoherence at least up to ${\it O}(\tau^{-3})$.

Considering Eqs.~(\ref{g0g1}), (\ref{gamma0}), (\ref{gamma1}), and (\ref{G2}), it is clear that we have obtained an effective decay rate, which is the leading order of Eq.~(\ref{gamma}) by Kofman and Kurizki. If one is interested to improve the accuracy of the approximated result in Eq.~(\ref{G2}), one then can refer to a different technique, e.g., that in Ref.~\cite{2018arXiv180401320L}. We also provide a comparison between our approximate result Eq.~(\ref{G2}) and a more quantitative approach in Appendix~\ref{cal} for a particular spectrum. However, one can see that a more rigorous calculation does not contradict the conclusion derived from Eq.~(\ref{G2}). Thus the criterion based on Eq.~(\ref{G2}) can be used to probe the occurrence of the QZE or the QAZE for many spectrums.

In Refs.~\cite{Chaudhry2014,Zhang2015,Chaudhry2016}, an alternative method based on the local properties of the decay rate is followed to characterize the QZE and the QAZE by numerical calculation. The QZE takes place when the effective decay rate decreases as the measurement interval becomes smaller, i.e., $d\gamma_{\rm eff}(\tau)/d\tau>0$. In contrast, the QAZE occurs when the effective decay rate $\gamma_{\rm eff}(\tau)$ increases as the measurement interval becomes smaller, i.e., $d\gamma_{\rm eff}(\tau)/d\tau<0$. One can check that our identification criterion of the QZE and the QAZE naturally covers the preceding result. In particular, when $G''(\Delta)<0$, $d\gamma_{\rm eff}(\tau)/d\tau\approx -8\pi G''(\Delta)/\tau^3>0$, which leads to the QZE; and when $G''(\Delta)>0$, $d\gamma_{\rm eff}(\tau)/d\tau<0$, which leads to the QAZE.

Equation~(\ref{G2}) is the main result of this work. We obtain a concise indicator to predict the quantum Zeno or the anti-Zeno effect, without the need to explicitly perform the involved integrals in the open-quantum-system dynamics. Under the energy-time uncertainty principle, $\Delta E\tau\sim 2\pi$, the successive measurements with a time interval $\tau$ will generate fluctuations of the system frequency $\Delta$ by an amount $2\pi\tau^{-1}$. It is the half-width of the main lobe of the filter function $F(\omega,\tau)$. As the energy uncertainty increases with reducing the measurement interval $\tau$, the excited atom decays via more and more channels around the frequency $\Delta$ and near-resonant modes take part in the process. Then Eq.~(\ref{G2}) actually provides the leading-order correction $\gamma_1(\tau)$ for the effective decay rate of the system by Fermi's golden rule, which only takes account the resonant contribution.

The shape of the SDF $G(\omega)$ around $\Delta$ therefore determines the lifetime of the system on the excited level. In particular, the second-order derivative of the spectral density function $G(\omega)$ at $\Delta$ will play an important role. When $G''(\Delta)<0$, the filter function $F(\omega,\tau)$ with a finite $\tau$ yields a smaller result for the integration in Eq.~(\ref{gamma}) than $F(\omega,\tau\rightarrow\infty)$ does. Thus under the measurement the effective transition between the TLS and the environment becomes weaker than that for the free decay. Frequent measurements thus inhibit the effective decay rate and result in the QZE. Otherwise when $G''(\Delta)>0$, frequent measurements enhance the interaction between the TLS and the environment, resulting in the QAZE.

\section{The QZE and the QAZE in various environments}\label{nonMar}

To apply our general criterion, in this section we examine the QZE or the QAZE in various structured environments by
\begin{equation}
\label{geff}
\tilde{\gamma}_{\rm eff}(\tau)=\gamma_0+\tilde{\gamma}_1(\tau)
\approx 2\pi G(\Delta)+\frac{4\pi}{\tau^2}G''(\Delta).
\end{equation}
For the sake of definiteness, here we suppose that the system frequency $\Delta$ is not extremely smaller than the cut-off frequency of the environment.

\subsection{Lorentzian spectral density}\label{lsd}

In the case of a two-level atom coupled to a cavity with high finesse mirrors~\cite{Lang1973,Gea-Banacloche1990,Ley1987}, the SDF can be approximated by a finite-band Lorentzian form
\begin{equation}\label{LF}
G_{L}(\omega)=\frac{D_0\Lambda^2}{(\omega-\omega_0)^2+\Lambda^2}.
\end{equation}
Here $\omega_0$ is the spectral center, $D_0$ is the spectral height and $\Lambda$ is the spectral halfwidth. The quantum open-system dynamics supported by the Lorentzian spectrum is exactly solvable within the RWA. One can easily obtain the exact decay rate~\cite{Kofman1996,Facchi2001,Koshino2005,Xu2014} (see Appendix~\ref{Lorent} for more details)
\begin{equation}
\label{gl}
\gamma(\tau)=-\frac{2}{\tau}\ln\left|\frac{a_+e^{-ia_-\tau}-a_-e^{-ia_+\tau}}{a_+-a_-}\right|
\end{equation}
where $a_\pm=[\Omega-i\Lambda\pm\sqrt{(\Omega-i\Lambda)^2+4\pi D_0\Lambda}]/2$, and $\Omega=\omega_0-\Delta$ is the detuning between the atom and the spectrum center. Alternatively, the effective decay rate based on the UT approach of Eq.~(\ref{gamma}) reads,
\begin{equation}
\label{LUT}
\gamma_{\rm eff}(\tau)=\gamma_0 \left[1+\frac{\sin\theta-\sin(\theta+\Omega\tau)e^{-\Lambda\tau}}{\Lambda\tau}\right],
\end{equation}
where $\theta$ satisfies $\sin\theta=(\Omega^2-\Lambda^2)/(\Omega^2+\Lambda^2)$ and $\cos\theta=2\Omega\Lambda/(\Omega^2+\Lambda^2)$, and $\gamma_0=2\pi G_L(\Delta)=2\pi D_0\Lambda^2/(\Omega^2+\Lambda^2)$. In addition, we can also obtain the approximated results from Eqs.~(\ref{G2}) and (\ref{g0g1}),
\begin{eqnarray}
G_L''(\Delta)&=&2D_0\Lambda^2\frac{3(\Delta-\omega_0)^2-\Lambda^2}{[(\Delta-\omega_0)^2+\Lambda^2]^3}, \\
\tilde{\gamma}_{\rm eff}(\tau)&=&\gamma_0\left[1+\frac{4(3\Omega^2-\Lambda^2)}{(\Omega^2+\Lambda^2)^2\tau^2}\right].
\label{L2}
\end{eqnarray}

\begin{figure}[htbp]
\centering
\includegraphics[width=0.5\textwidth]{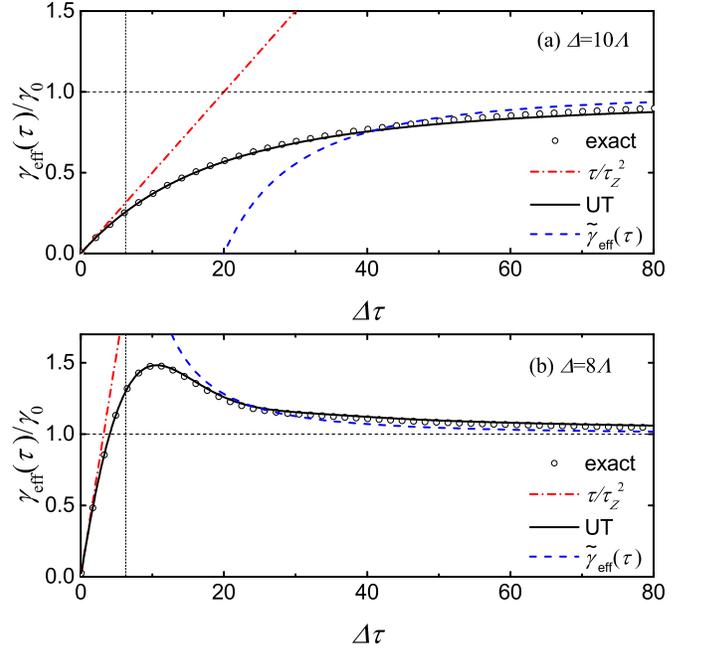}
\caption{(Color online) The effective decay rate vs $\Delta\tau$ for the Lorentzian spectrum of (a) $\Delta=10\Lambda$ and (b) $\Delta=8\Lambda$, respectively. Other parameters are chosen as $\omega_0=10\Lambda$ and $D_0=0.01\Lambda$. The line with open circles represents the exact decay rate given by Eq.~(\ref{gl}). Here and in the following figures, the red dot-dashed line is the result of the linear approximation with the Zeno time, which is valid only for extremely short measurement intervals; the black solid line is effective decay rate based on the UT approach; and the blue dashed line is our approximated result $\tilde{\gamma}_{\rm eff}$. The horizontal and vertical dashed lines are used to depict $\gamma_{\rm eff}=\gamma_0$ and mark the position of $\Delta\tau=2\pi$, respectively. }
\label{Lorentz}
\end{figure}

The above three effective decay rates for the Lorentzian spectral function are displayed in Fig.~\ref{Lorentz}. Here we restrict our comparison to the weak-coupling regime, where $D_0=0.01\Lambda$ and $\omega_0=10\Lambda$. We also plot the linear approximation $\gamma(\tau)\approx\tau/\tau_z^2$ for the short-time limit (the details about the Zeno time $\tau_z$ are provided in Sec.~\ref{Zt}).

The curve based on UT approach is shown to fit well with the exact solution in Fig.~\ref{Lorentz}. This agreement ensures that in the weak-coupling regime, the UT approach is a good approximation to the exact result. It is then used as a benchmark in the following spectrums with no analytical results. When we consider the effective decay rate for large times $\Delta\tau\gtrsim 2\pi$, which is accessible for practical measurement, our approximated result in Sec.~\ref{concavity} is in a good agreement with the exact solution and that from the UT approach in both quality and quantity. The accuracy of our result is enhanced as the measurement spacing $\tau$ increases. The omitted part in Eq.~(\ref{G2}), which is in the order of ${\it O}(\tau^{-3})$, will become even smaller in this situation. In general, a SDF that does {\em not} grow drastically when the frequency is beyond $\Delta$ is helpful to match our approximation with the exact results since we take a loose integration domain to obtain Eq.~(\ref{G2}).

These curves show that if the atomic frequency is around the spectral center [see Fig.~\ref{Lorentz}(a)], where $\Omega=\omega_0-\Delta=0$ and then $G_L''(\Delta)<0$, $\ti{\gamma}_{\rm eff}$ is always smaller than the free decay rate $\gamma_0$. The decay gets more suppressed as the measurements become more frequent. Thus the QZE is observable in almost the whole domain of the measurement intervals. On the other hand, if the detuning frequency $\Omega$ is considerably larger than the central frequency [see Fig.~\ref{Lorentz}(b) where $\Omega=\omega_0-\Delta=2\Lambda$], then $G_L''(\Delta)>0$ and the repeated measurements can speed up the decay rate, i.e., $\ti{\gamma}_{\rm eff}$ may become larger than the free decay rate, giving rise to the QAZE. These results indicate that open systems under an environment with the Lorentzian spectrum are all good testbeds to observe the quantum Zeno or the anti-Zeno effect. The numerical result demonstrates that our simple and explicit criterion can be easily applied to predict the QZE or the QAZE for large times of decay for the case of Lorentzian form spectrum.

\subsection{Hydrogenlike spectral density}

\begin{figure}[htbp]
\centering
\includegraphics[width=0.5\textwidth]{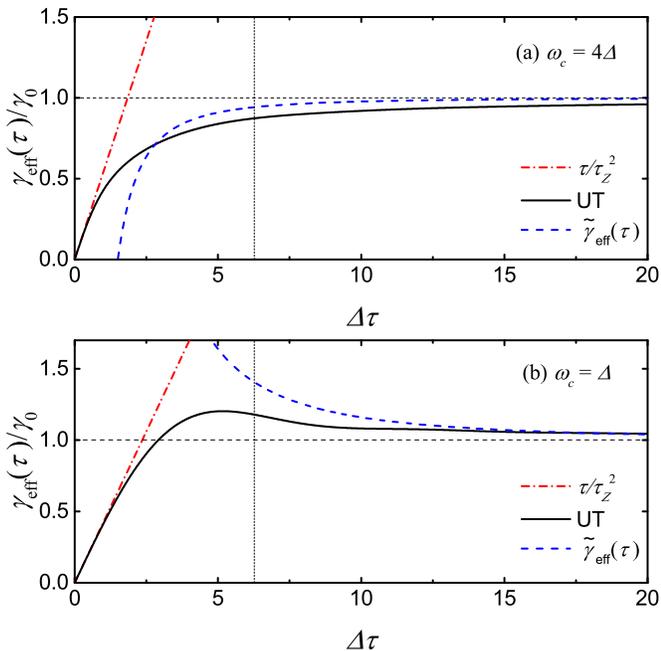}
\caption{(Color online) The effective decay rate vs $\Delta\tau$ for the hydrogenlike spectral density function with (a) $\omega_c=4\Delta$ and (b) $\omega_c=\Delta$. }
\label{Hydrogen}
\end{figure}

Next we consider a hydrogenlike spectral density function
\begin{equation}
G_h(\omega)=\frac{\eta\omega}{\left[1+\left(\frac{\omega}{\omega_c}\right)^2\right]^4},
\end{equation}
as given in Ref.~\cite{Moses1973,Facchi1998,Zheng2008,Li2009,Ai2010}, where $\eta$ is the dimensionless coupling strength and $\omega_c$ is the cut-off frequency. The numerical calculations of the effective decay rate in Eqs.~(\ref{gamma}) and (\ref{g0g1}) are shown in Fig.~\ref{Hydrogen}. As it is hard to obtain an exact analytical expression for the evolution of the full system, we compare our approximated results in Eq.~(\ref{g0g1}) with those from Eq.~(\ref{gamma}) by the UT approach. In this case, the second derivative of the SDF is
\begin{equation}\label{Gh}
G_h''(\Delta)=8\eta\Delta\omega_c^8\frac{7\Delta^2-3\omega_c^2}{(\Delta^2+\omega_c^2)^6}.
\end{equation}
The two results fit well with each other. Here we choose $\omega_c=4\Delta$ and $\omega_c=\Delta$ in Figs.~\ref{Hydrogen}(a) and \ref{Hydrogen}(b) respectively to show the occurrence of both the QZE and QAZE, respectively. When $\Delta=\omega_c/4$ or smaller, the QZE always dominates and the QAZE never occurs (remember that the case where $\Delta\ll\omega_c$ is not covered here). The boundary of these two effects can be approximately estimated by Eq.~(\ref{Gh}), i.e., $\om_c\approx\sqrt{7/3}\Delta\approx1.53\Delta$. In fact, the QZE is a general consequence for a natural hydrogenlike atom subject to the vacuum electromagnetic fields via the electric dipole transition $2P-1S$. For a natural atom, the coupling constant $\eta\propto Z^2\alpha^3$, the system energy $\Delta\propto Z^2\alpha^2$, and the cut-off frequency $\omega_c\propto Z\alpha$, where $Z$ is the nuclear charge number and $\alpha=1/137$ is the fine-structure constant, then the ratio $\Delta/\omega_c$ scales with $Z\alpha$~\cite{Facchi1998}. The nuclear charge number $Z$ has to be above a hundred, for this ratio to approach unity: the QAZE is very hard to achieve for (electric dipole transitions in) natural atoms. Only if the transition frequency of an artificial atom is close to the cut-off frequency $\omega_c$, does the second derivative $G_h''(\Delta)$ then become positive. The QAZE thus appears and determines the decay pattern.

A counter fact should be noted here that if the ratio $\Delta/\omega_c$ is extremely small, such as $\Delta/\omega_c=1/550$ for the hydrogen atom, the QAZE occurs~\cite{Ai2010}. In this extreme case, the contribution from the integration interval beyond $2\pi/\tau$ becomes significant to $\gamma_1(\tau)$ in Eq.~(\ref{gamma1}), so that Eq.~(\ref{G2}) now fails to be a good approximation to Eq.~(\ref{gamma1}), and then our criterion loses its ability to predict this anti-Zeno phenomenon.

\subsection{Power-law spectral density}\label{powerlaw}

The power-law spectral density functions can be expressed by a general function~\cite{Leggett1987}
\begin{equation}
G_{p}(\omega)=A\omega_c^{1-s}\omega^se^{-\omega/\omega_c},
\end{equation}
where $A$ is a dimensionless coupling strength and $\omega_c$ is the cut-off frequency. Here we choose an exponential cut-off form. The bath exponent $s$ categorizes the environments into the super-Ohmic ($s>1$), Ohmic ($s=1$), and sub-Ohmic ($s<1$) types, respectively. We set $\omega_c=10\Delta$ for the data presented below and constrain the calculation to the weak-coupling regime, where $A$ is small enough to ensure the validity of the UT approach.

\begin{figure}[htbp]
\centering
\includegraphics[width=0.5\textwidth]{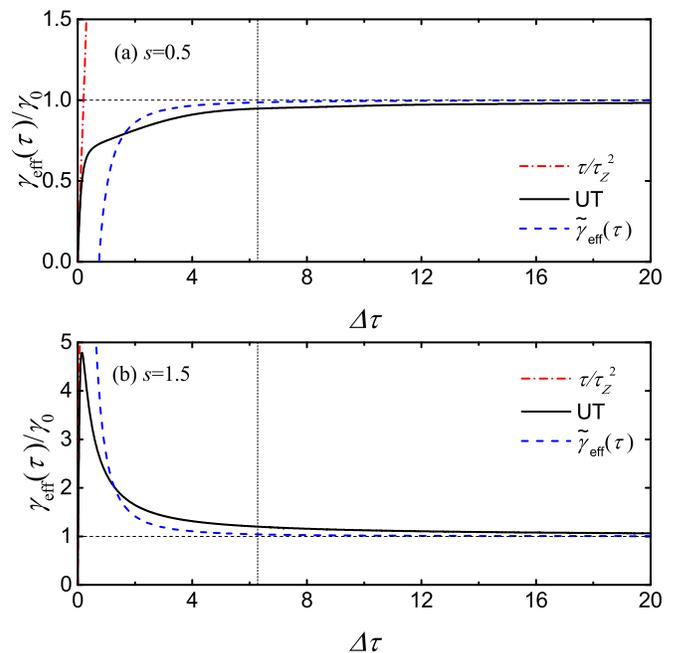}
\caption{(Color online) The effective decay rate vs $\Delta\tau$ for the power-law spectral density functions with (a) $s=0.5$ and (b) $s=1.5$. The cut-off frequency is chosen as $\omega_c=10\Delta$. }
\label{power}
\end{figure}

The effective decay $\gamma_{\rm eff}/\gamma_0$ versus the time-spacing of measurement $\tau$ is plotted in Figs.~\ref{power}(a) and \ref{power}(b) for the sub-Ohmic and super-Ohmic cases, respectively. These curves again show that our approximation agrees well with the result of the UT approach in quality for a large domain of $\tau$. Under a sub-Ohmic bath, the effective decay rate of the system gradually increases to $\gamma_0$. In contrast, under a super-Ohmic bath, the effective decay rate gradually goes down to the free decay rate $\gamma_0$. In fact, the second derivative of the power-law spectral density function with a large cutoff $\omega_c$ reads
\begin{equation}
G_p''(\Delta)\approx s(s-1)A\omega_c^{1-s}\Delta^{s-2}.
\end{equation}
The criterion is obviously determined by $s$. When $0<s<1$, $G_p''(\Delta)<0$, which leads to the QZE [see Fig.~\ref{power}(a)]. In contrast, when $s>1$, $G_p''(\Delta)>0$, which leads to the QAZE [see Fig.~\ref{power}(b)]. In other words, the QZE always dominates for sub-Ohmic baths and the QAZE always dominates for super-Ohmic baths.

However, our criterion is no longer valid for Ohmic baths, where $G''_p(\Delta)\approx 0$. In this case, the contribution in the order of $\tau^{-1}$ would play an important role in the integral in Eq.~(\ref{g0g1}). As the second derivative of $G_p(\Delta)$ is near zero, the only feature one can capture is that the effective decay rate approaches the free decay rate in a much faster way than those in the sub-Ohmic and super-Ohmic cases. When $s>1$, especially when $s>2$, the center of gravity of $G(\omega)$, which locates at $s\om_c$, is remarkably far away from the energy difference $\Delta$ of the TLS, seemingly putting our approximate results in question. We can consider a more rigorous calculation, which is performed in Appendix~\ref{cal}. The conclusion remains unchanged that a super-Ohmic bath with $s>1$ leads to the QAZE. Therefore, our criterion serves as a sufficient condition to identify the Q(A)ZE in quality.

\section{Discussion}\label{dissc}

\subsection{Weak-coupling regime}

It should be emphasized that our criterion as well as the previous UT approach is appropriate for situations when the system-environment coupling is weak. It is equivalent to the first-order time-dependent perturbation theory. In the weak-coupling regime, the correlation between the TLS and environment is so weak that the state of the environment does not change significantly during the repetitive measurements. Thus one can assume that the total system collapses to its initial state after each measurement. Moreover, the decay pattern of an open-quantum system is often nonexponential~\cite{Beau2017}. One can always use the effective decay rate to describe the system evolution in a moderate time scale under the weak-coupling approximation. However, when $\tau$ becomes extremely small, the measurement-induced broadening of the excited state would reduce the effective energy splitting (the original splitting $\Delta$ plus the fluctuation induced by frequent measurements) between the excited state and the ground state of the two-level system. Consequently, the ratio of system-environment coupling strength and effective energy splitting will go out of the weak-coupling regime. The rotating-wave approximation will subsequently fail. In particular, the energy spread may destroy the observed system, leading to the creation of new particles in the worst case, which is incompatible with the two-level approximation~\cite{Kofman2000}. Thus a smaller $\tau$ than $2\pi/\Delta$ might be inconsistent with the starting point of our model~(\ref{H0}) and then be out of the reach of our criterion.

Additionally, in the weak-coupling regime, Eq.~(\ref{gamma}) shows that the effective decay rate increases with the coupling strength. However, the ratio of $\gamma_{\rm eff}/\gamma_0$ is independent of the coupling strength. Namely, proper scaling of the coupling strength has no effect on the occurrence of the QZE or the QAZE. By contrast, in the strong-coupling regime, the coupling strength actually decreases the effective decay rate and the ratio of $\gamma_{\rm eff}/\gamma_0$ is closely relevant to the coupling strength~\cite{Chaudhry2017}. Thus our criterion is not appropriate in the strong-coupling regime.

\subsection{Beyond the rotating-wave approximation}

It is well known that the RWA offers a good approximation in the weak-coupling limit. To consider the effect of the counter-rotating terms in the Hamiltonian, Eq.~(\ref{H0}) is modified as
\begin{eqnarray} \label{nRWA}
\mathcal{H}&=&H_0+\mathcal{H}_{I}, \\ \non
H_0&=&\Delta\sigma^+\sigma^-+\sum_k\omega_k b_k^{\dag}b_k,\\ \non
\mathcal{H}_{I}&=&\sum_k\left[g_k^*(b_k+b_k^\dag)\sigma^++{\rm H.c.}\right].
\end{eqnarray}
A generalized version of the Fr\"{o}hlich-Nakajima transformation $\exp(S)$ can be employed to eliminate the high-frequency terms in the effective Hamiltonian~\cite{Ai2010}. When the anti-Hermitian operator $S$ is chosen as
\begin{equation}\label{HS}
S=\sum_k \frac{1}{\omega_k+\Delta}(g_k^*b_k^\dag\sigma^+-g_kb_k\sigma^-),
\end{equation}
The effective Hamiltonian $H_{\rm eff}=\exp(S)\mathcal{H}\exp(-S)$ is approximated as
\begin{equation}
H_{\rm eff}=\Delta_1\sigma^+\sigma^-+\sum_k\omega_kb_k^\dag b_k+\sum_k\left(g_k^*b_k\sigma^++g_kb_k^{\dag}\sigma^-\right),
\end{equation}
up to the second order. Here the modified level spacing for the TLS, which includes the Lamb shift, reads
\begin{equation}
\Delta_1=\Delta+\sum_k\frac{|g_k|^2}{\omega_k+\Delta}.
\end{equation}
For details, one can refer to Appendix~\ref{WRWA}. Due to the special unitary transformation $\exp(S)$, the initial states before and after the transformation are identical, and the effective Hamiltonian has the same form as the Hamiltonian with the RWA. Therefore, we can obtain the effective decay rate as before by merely replacing $\Delta$ with $\Delta_1$. In the weak-coupling regime, $\Delta_1\approx\Delta$, the effect of counter-rotating terms thus can be neglected in our criterion.

\subsection{Zeno time and practical measurement interval}\label{Zt}

When the time interval $\tau$ is extremely short, the survival probability is quadratic in $\tau$,
\begin{equation}\label{st}
p(\tau)=1-\frac{\tau^2}{\tau_Z^2},
\end{equation}
which means that the Zeno effect always appears for a sufficiently short $\tau$. The Zeno time $\tau_Z$~\cite{Facchi1998,Facchi2001,Zheng2008} defines the time of the Zeno effect and can be evaluated by
\begin{equation}
\tau_Z=\lim_{\tau\to 0}\left(\frac{\gamma(\tau)}{\tau}\right)^{-1/2}=
\left(\int_0^\infty d\omega G(\omega)\right)^{-1/2}.
\end{equation}
Then we can obtain another effective decay rate within the linear approximation,
\begin{equation}
\gamma(\tau)\approx \frac{\tau}{\tau_Z^2}.
\end{equation}
The linear-approximation results for different cases have been shown in Figs.~\ref{Lorentz}, \ref{Hydrogen}, and \ref{power} by red dot-dashed lines. It is remarkable that the QZE can be observed in any open-quantum-system when $\tau$ is sufficiently small, but the QAZE does not necessarily occur. The decay behaviors shown in section~\ref{nonMar} can be classified into two situations. In Figs.~\ref{Lorentz}-\ref{power}(a), the effective decay rates obtained by the UT approach gradually increase towards $\gamma_0$ as the measurement spacing increases, which demonstrates the Zeno effect. This phenomenon can be predicted by our criterion at least in a certain domain. In Figs.~\ref{Lorentz}-\ref{power}(b), the effective decay rates obtained by the UT approach increase first and then decrease, which shows a transition from the Zeno to the anti-Zeno effect. However, our criterion can capture merely the latter phenomenon but not the whole pattern (see the blue dashed lines). Notice that the Zeno time $\tau_Z$ is extremely small and hard to access in practice, e.g., it is about $3.59\times10^{-15}$ s for the $2P-1S$ transition of the hydrogen atom~\cite{Facchi1998}. Therefore, it is experimentally challenging to observe the initial quadratic behavior and enforce any sequences of measurements into such a small timescale.

Thus a proper measurement spacing, which is much larger than the Zeno time is more meaningful to multiple measurements and then also to probe the QZE or the QAZE phenomenon. This proper spacing defined as $\tau\gtrsim 2\pi/\Delta$ in our work, is based on three considerations. The first one is that one should ensure the integrity of the main lobe of the filter function $F(\omega,\tau)$ in the integration to obtain a measurement-modified effective decay rate~(\ref{gamma}). The second one is that $\tau\gtrsim 2\pi/\Delta$ can ensure the ratio of system-environment coupling strength and effective energy splitting is in the weak-coupling regime, which is consistent with the rotating-wave approximation in our model. The last one is that it is difficult to enforce a sequence of measurements with even smaller timescale in practical experiments.

In some cases, although our criterion fails to directly predict the Zeno effect with a smaller $\tau$, it predicts an anti-Zeno effect with a larger $\tau$. Then, logically, there is at least one transition which separates the whole domain of $\tau$ into a Zeno regime and an anti-Zeno regime. But it is beyond our reach.

\subsection{Application ranges of our criterion}

In this work, we apply two approximations to deduce our Q(A)ZE criterion. The first one is to use the main lobe integration domain instead of the whole frequency domain from $0$ to $\infty$, which requires that the system energy $\Delta$ cannot be extremely smaller than both the center of gravity of $G(\omega)$ and the cut-off frequency $\om_c$ (if it exists). The second one is to consider the effect from the SDF $G(\omega)$ around $\Delta$ up to its second derivative with respect to frequency, which requires that $G''(\Delta)$ cannot be extremely small in magnitude compared to the higher orders of derivative of $G(\omega)$. Otherwise our criterion will be overwhelmed either by the contributions from the integration outside of the main lobe of the filter function or the higher derivatives of $G(\omega)$. We need to check the two general principles in the particular spectra to reveal the application range of our criterion.

\begin{figure}[htbp]
\centering
\includegraphics[width=0.5\textwidth]{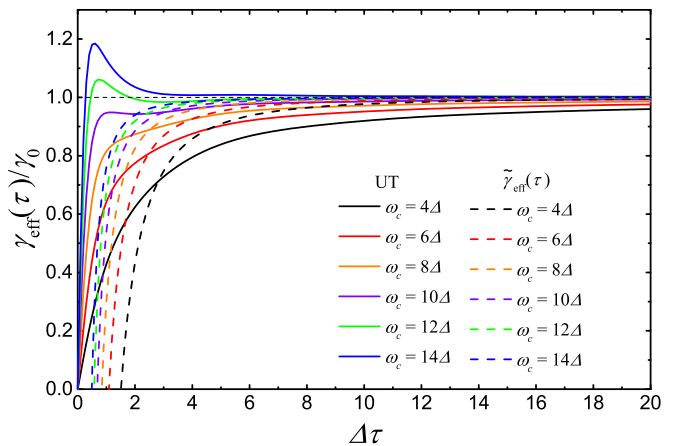}
\caption{(Color online) The effective decay rate vs $\Delta\tau$ for different cut-off frequencies for the Hydrogenlike spectrum. }
\label{H1}
\end{figure}

\begin{figure}[htbp]
\centering
\includegraphics[width=0.5\textwidth]{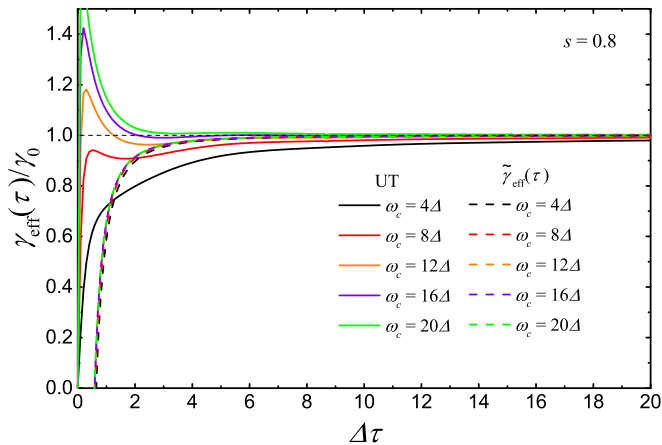}
\caption{(Color online) The effective decay rate vs $\Delta\tau$ for different cut-off frequencies for the sub-Ohmic bath with $s=0.8$. }
\label{p1}
\end{figure}

For the Lorentzian spectrum, the condition of the QZE with a sufficiently large measurement time-spacing can be deduced as about $|\Omega|<\Lambda$ by Eq.~(\ref{LUT})~\cite{Koshino2005}, while our criterion yields about $1.73|\Omega|<\Lambda$ by Eq.~(\ref{L2}). Roughly these two results match with each other. For the hydrogenlike spectrum, we plot the effective decay rate versus $\Delta\tau$ for different cut-off frequencies in Fig.~\ref{H1}. According to Eq.~(\ref{Gh}), the QZE occurs when the cut-off frequency $\om_c$ is larger than about $1.53\De$. As seen from the UT approach, the QAZE may occur when the cut-off frequency is enhanced over about $12\De$, which is beyond our prediction. This is due to the fact that in this condition, the system energy is faraway from the cut-off frequency, and then contributions from the minor lobes of the filter function overwhelm those from the main lobe. The same reason may lead to the QAZE occurring for sub-Ohmic baths with $s<1$, which contradicts our conclusion that the QZE always dominates for the sub-Ohmic baths in Sec.~\ref{powerlaw}. In Fig.~\ref{p1}, we plot the effective decay rate versus $\Delta\tau$ for different cut-off frequencies with $s=0.8$. In the practical regimes of $\tau$ (omitting the intervals which are too small to realize in experiments), our criterion works well unless the cut-off frequency $\omega_c$ is enhanced to about $20\Delta$. A smaller $s$ can give rise to an even larger upper bound for $\om_c$.

\section{Conclusion}\label{concl}

In summary, we have investigated an explicit criterion about the Zeno or the anti-Zeno effect in a two-level system coupled to a dissipative environment. Based on the conventional unitary transformation approach, we have found that with a moderate time-spacing, which is accessible in practical experiments, the Zeno or the anti-Zeno effect is closely related to the second derivative of the spectrum $G''(\Delta)$ with respect to the energy difference of the two-level system. In particular, $G''(\Delta)<0$ and $G''(\Delta)>0$ indicate the QZE and the QAZE, respectively. As the initial quadratic behavior of the survival probability in Eq.~(\ref{st}) is hardly observable in experiments, more and more theoretical and experimental attention is paid to the timescales allowing multiple measurements in practice. Our work connects the phenomena of the QZE or the QAZE and the convexity of the spectrum in this time regime and serves as a simple criterion for the Q(A)ZE in quality, but might not in quantity.

Our criterion helps one to conveniently identify the QZE or the QAZE pattern in the open-system dynamics under various spectral density functions. In the Lorentzian spectrum, we found that if the atomic frequency is around the spectral center, the decay rate of the system then gets more suppressed when the measurements become more frequent, i.e., the QZE occurs. In contrast, if the detuning between the atom and the spectral center is sufficiently large, then the QAZE occurs. In the hydrogenlike spectrum, the QZE dominates for almost all the natural atoms where the transition frequency $\Delta$ is considerably less than the cut-off frequency $\omega_c$, and the QAZE occurs only if $\Delta\approx\omega_c$. In the power-law spectrum, we have found that the QZE and the QAZE appear in the sub-Ohmic and super-Ohmic bath, respectively.

It should be noted that our criterion is valid in the weak-coupling regime, where the counter-rotating terms in the interaction Hamiltonian can be safely neglected. Our work is of interest to the measurement and control of open-quantum systems.

\section*{Acknowledgments}
We acknowledge grant support from the National Science Foundation of China (Grant No. 11575071), the Fundamental Research Funds for the Central Universities, and Zhejiang Provincial Natural Science Foundation of China under Grant No. LD18A040001.

\appendix

\section{General expression for the effective decay rate}

In the interaction picture with respect to $H_0=\Delta\sigma^+\sigma^-+ \sum_k\omega_k b_k^{\dag}b_k$, the full Hamiltonian~(\ref{H0}) is rewritten as,
\begin{equation}\label{HI}
\begin{aligned}
&H_I=U_0(H-H_0)U_0^{-1}\\
&=\sum_k\left(g_k^*b_k\sigma^+e^{i(\Delta-\omega_k)t} +g_kb_k^{\dag}\sigma^-e^{-i(\Delta-\omega_k)t}\right),
\end{aligned}
\end{equation}
upon the unitary transformation represented by $U_0=\exp(iH_0t)$. Starting from the state $|\psi(0)\rangle=|e,0\rangle$, e.g., the two-level system is in the excited state and the environment is in the vacuum state, the entire state of system and environment at time $t$ can be written as
\begin{equation}\label{psi}
|\psi(t)\rangle=\alpha(t)|e,0\rangle +\sum_kc_k(t)|g,k\rangle,
\end{equation}
where $|g,k\rangle$ means that the state of the TLS is in the ground state and the $k$-th mode of the environment is in the first excited state. The total number of excitons is conserved under the action of the full Hamiltonian $H_I$.

\subsection{The unitary-transformation approach}\label{UT}

From the Schr\"odinger equation $\partial_t|\psi(t)\rangle=-i H_I|\psi(t)\rangle$, one can obtain the following equations
\begin{eqnarray}\label{dalpha}
\dot{\alpha}(t)&=&-i\sum_k g_k^*c_k(t)e^{i(\Delta-\omega_k)t},
\\ \label{dck}
\dot{c}_k(t)&=&-ig_k\alpha(t)e^{-i(\Delta-\omega_k)t}.
\end{eqnarray}
Formally integrating Eq.~(\ref{dck}) yields
\begin{equation}\label{ck}
c_k(t)=-i g_k\int_0^t dt'\alpha(t')e^{-i(\Delta-\omega_k)t'}.
\end{equation}
Inserting Eq.~(\ref{ck}) into Eq.~(\ref{dalpha}), we obtain an exact integro-differential equation
\begin{equation}\label{alpha}
\dot{\alpha}(t)=-\int_0^tdt'e^{i\Delta(t-t')}\Phi(t-t')\alpha(t'),
\end{equation}
where $\Phi(t)=\sum_k|g_k|^2e^{-i\omega_k t}$. Constrained by a sufficiently short time $t$, such that $\alpha(t)\simeq\alpha(0)=1$ yet long enough to allow the rotating-wave approximation, Eq.~(\ref{alpha}) yields
\begin{equation}\label{alpha1}
\alpha(t)\simeq 1-\mathcal{I}(t),
\end{equation}
where
\begin{equation}
\begin{aligned}
\mathcal{I}(t)=&\int_0^t dt'\int_0^{t'}dt''e^{i\Delta(t'-t'')}\Phi(t'-t'')\\
=&\int_0^t dt'(t-t')\Phi(t')e^{i\Delta t'}.
\end{aligned}
\end{equation}
Here $\alpha(t)$ is much more reliable than a commonly accepted equation, Eq.~(\ref{st}), as it takes into account all the powers of $t$. We can then find the survival probability that the TLS is still in its excited state is $p(t)=|\alpha(t)|^2$.

Equation~(\ref{psi}) implies that if we detect the TLS to be excited by a device, then we can immediately deduce that the environment is still in the vacuum state. On the other hand, if we make the null-result measurements on the environment, then we can confirm that the TLS does not decay. In this sense, the entire state keeps being reset upon each measurement. After $N$ measurements, each performed after a time interval $\tau$, the survival probability of the excited state is
\begin{equation}\label{s}
p(t=N\tau)=p^N(\tau)=|\alpha(\tau)|^{2N}\equiv e^{-\gamma(\tau)t},
\end{equation}
where $\gamma(\tau)$ is the decay rate. With the UT approach, the effective decay rate is given by
\begin{equation}
\begin{aligned}
\gamma_{\rm eff}(\tau)
=&-\frac{1}{\tau}\ln |a(\tau)|^2\\
\approx &-\frac{1}{\tau}\ln\big(1-2{\rm Re}[\mathcal{I}(\tau)]+|\mathcal{I}(\tau)|^2\big)\\
\approx &\frac{2}{\tau}{\rm Re}[\mathcal{I}(\tau)]=2{\rm Re}\left[\int_0^\infty dt f(t)\Phi(t)\right].
\end{aligned}
\end{equation}
Here the measurement effects are accounted for by $f(t)=(1-t/\tau)e^{i\Delta t}\theta(\tau-t)$, where the step function $\theta(t)$ is $1$ for $t>0$ and $0$ for $t<0$. We can recast the expression by applying the spectral density function $G(\omega)=\sum_k|g_k|^2\delta(\omega-\omega_k)$, which is the Fourier transformation of $\Phi(t)$. Then the effective decay rate yields
\begin{equation}\label{reff}
\begin{aligned}
\gamma_{\rm eff}(\tau)=&2{\rm Re}\left[\int_0^\tau dt \left(1-\frac{t}{\tau}\right)e^{i\Delta t}\int_0^\infty d\omega G(\omega)e^{-i\omega t}\right]\\ =&2\pi\int_0^\infty d\omega G(\omega)F(\omega,\tau),
\end{aligned}
\end{equation}
where
\begin{equation}
F(\omega,\tau)=\frac{\tau}{2\pi}{\rm sinc}^2\left[\frac{(\omega-\Delta)\tau}{2}\right],
\end{equation}
serves as a filter function.

\subsection{The first-order perturbation theory}\label{first}

As for the perturbation theory, it is straightforward to derive the expansion for the evolution operator:
\begin{eqnarray} \non
&& \mathcal{T}_{\leftarrow}\left\{\exp\left[-i\int_0^tdt'H_I(t')\right]\right\}=I+(-i)\int_0^tdt'H_I(t')
\\ &+&(-i)^2\int_0^tdt'\int_0^{t'}dt''H_I(t')H_I(t'')+\cdots,
\end{eqnarray}
where $H_I(t)$ is given by Eq.~(\ref{HI}). Now we calculate the decay amplitude $c_k(t)$ defined in Eq.~(\ref{psi}) within the leading-order perturbation, while omitting the second order and even higher orders of $H_I(t)$. Thus $c_k(t)$ is reduced to the following form:
\begin{equation}
\begin{aligned}
c_k(t)\approx &-ig_k^*\int_0^tdt'e^{-i(\Delta-\omega_k)t'}\\
=&-ig_k^*te^{-i(\Delta-\omega_k)t/2}{\rm sinc}[(\Delta-\omega_k)t/2].
\end{aligned}
\end{equation}
We can then find the survival probability is
\begin{eqnarray} \non
p(t)&=&1-\sum_k|c_k(t)|^2 \\ &\approx& 1-t^2\sum_k|g_k|^2{\rm sinc}^2[(\Delta-\omega_k)t/2].
\end{eqnarray}
If the state is reset to the atomic excited state after every confirmation of survival, then the survival probability of the excited atom is $p(t=n\tau)=p^n(\tau)=[p(\tau)]^{t/\tau}$. Therefore, the effective decay rate under such repeated measurements is given by
\begin{equation}
\gamma_{\rm eff}(\tau)=-\frac{1}{\tau}\ln p(\tau)\approx\tau\sum_k|g_k|^2{\rm sinc}^2[(\Delta-\omega_k)\tau/2],
\end{equation}
as a function of the measurement interval $\tau$. It coincides with Eq.~(\ref{reff}).

\section{The evaluation of $\gamma_1(\tau)$ in Eq.~(\ref{g0g1})}\label{cal}

The measurement-modified effective decay rate in Eq.~(\ref{gamma}) can be decomposed into two parts:
\begin{equation}
\begin{aligned}
\gamma_{\rm eff}(\tau)=&2\pi\int_0^{\infty}d\omega G(\omega)F(\omega,\tau)\\
=&2\pi\int_0^{\infty}d\omega [G(\omega)-G(\Delta)+G(\Delta)]F(\omega,\tau)\\
=&2\pi G(\Delta)\int_0^{\infty}d\omega F(\omega,\tau)\\
&+2\pi\int_0^{\infty}d\omega [G(\omega)-G(\Delta)]F(\omega,\tau)\\
=&\gamma_0+\gamma_1(\tau).
\end{aligned}
\end{equation}
Here $\gamma_0$ is a measurement-independent part,
\begin{equation}
\begin{aligned}
\gamma_0=&2\pi G(\Delta)\int_0^{\infty}d\omega F(\omega,\tau) \\
= & G(\Delta)\int_{-\Delta}^{\infty} \tau{\rm sinc}^2\left(\frac{x\tau}{2}\right)dx\\
\approx &G(\Delta)\int_{-\infty}^{\infty} \tau{\rm sinc}^2\left(\frac{x\tau}{2}\right)dx
=2\pi G(\Delta),
\end{aligned}
\end{equation}
where we change the integration variable to $x=\omega-\Delta$ and then expand the lower bound of the integration from $-\Delta$ to $-\infty$ to obtain $\gamma_0=2\pi G(\Delta)$ by Fermi's golden rule. $\gamma_1(\tau)$ is a measurement-dependent part,
\begin{equation}
\gamma_1(\tau)=2\pi\int_0^{\infty}d\omega [G(\omega)-G(\Delta)]F(\omega,\tau)
\end{equation}
and it is obvious that the sign of $\gamma_1(\tau)$ determines the occurrence of the QZE or the QAZE .

For the filter function $F(\omega,\tau)$, the main lobe which ranges from $\Delta-2\pi/\tau$ to $\Delta+2\pi/\tau$, covers about $90.3\%\left(=\int_{\Delta-2\pi/\tau}^{\Delta+2\pi/\tau}F(\omega,\tau)d\omega/ \int_{-\infty}^\infty F(\omega,\tau)d\omega\right)$ of the whole domain of frequency. When the difference between $\Delta$ and the center of gravity of $G(\omega)$ is not extremely large, it is reasonable to replace the integration domain $[0,\infty)$ by $[\Delta-2\pi/\tau,\Delta+2\pi/\tau]$, and ignore the contributions from the integration domain involving all the minor lobes. We then expand $G(\omega)$ around $\Delta$ by the Taylor's series, $G(\omega)=\sum_{n=0}G^{(n)}(\Delta)(\omega-\Delta)^n/n!$, to obtain
\begin{equation}
\begin{aligned}
\tilde{\gamma}_1(\tau)=&2\pi\int_{\Delta-2\pi/\tau}^{\Delta+2\pi/\tau}d\omega [G(\omega)-G(\Delta)]F(\omega,\tau)\\
=&\frac{2}{\tau}\int_{\Delta-2\pi/\tau}^{\Delta+2\pi/\tau}d\omega \left[\sum_{n=0}\frac{G^{(n)}(\Delta)}{n!}(\omega-\Delta)^n-G(\Delta)\right]\\ &\times\frac{1-\cos[(\omega-\Delta)\tau]}{(\omega-\Delta)^2}\\
=&\frac{2}{\tau}\int_{\Delta-2\pi/\tau}^{\Delta+2\pi/\tau}d\omega \sum_{n=1}\frac{G^{(n)}(\Delta)}{n!}(\omega-\Delta)^{n-2}\\
&\times\{1-\cos[(\omega-\Delta)\tau]\}\\
=&\frac{4\pi}{\tau^2}G''(\Delta)+{\it O}(\tau^{-3}).
\end{aligned}
\end{equation}
This is the result of Eq.~(\ref{G2}) in the main text.

To obtain a more rigorous result, one should consider the whole domain from $0$ to $\infty$. The total integral can be written as
\begin{equation}
\gamma_1(\tau)=\tilde{\gamma}_1(\tau)+\gamma_1^+(\tau)+\gamma_1^-(\tau),
\end{equation}
where
\begin{equation}
\gamma_1^+(\tau)=2\pi\int_{\Delta+2\pi/\tau}^\infty d\omega [G(\omega)-G(\Delta)]F(\omega,\tau),
\end{equation}
and
\begin{equation}
\gamma_1^-(\tau)=2\pi\int_0^{\Delta-2\pi/\tau} d\omega [G(\omega)-G(\Delta)]F(\omega,\tau).
\end{equation}
In the integration over the minor lobes of $F(\omega,\tau)$, we assume the value of the integral is insensitive to the rapidly oscillating behavior of $F(\omega,\tau)$, and we can simply replace the square sine function by its mean value $1/2$~\cite{2018arXiv180401320L},
\begin{equation}
\begin{aligned}
F(\omega,\tau)=&\frac{\tau}{2\pi}\frac{\sin^2[(\omega-\Delta)\tau/2]}{[(\omega-\Delta)\tau/2]^2}
\approx\frac{\tau}{2\pi}\frac{\frac{1}{2}}{[(\omega-\Delta)\tau/2]^2}\\
=&\frac{1}{\pi(\omega-\Delta)^2\tau}.
\end{aligned}
\end{equation}
Consequently we get
\begin{equation}
\gamma_1^+(\tau)\approx \frac{2}{\tau}\int_{\Delta+2\pi/\tau}^\infty d\omega \frac{G(\omega)-G(\Delta)}{(\omega-\Delta)^2},
\end{equation}
and
\begin{equation}
\gamma_1^-(\tau)\approx \frac{2}{\tau}\int_0^{\Delta-2\pi/\tau} d\omega \frac{G(\omega)-G(\Delta)}{(\omega-\Delta)^2}.
\end{equation}
Thus, when the wing of $G(\omega)$ grows slower than $\omega$, the integrand rapidly falls off and the integration for $\gamma_1^+(\tau)$ and $\gamma_1^-(\tau)$ can be ignored.

When $G(\omega)$ grows faster than $\omega$, such as the power-law spectral density function $G_p(\omega)=A\omega_c^{1-s}\omega^s e^{-\omega/\omega_c}$ with $s>1$, $\gamma_1^+(\tau)$ and $\gamma_1^-(\tau)$ should be considered.
Changing the integration variable to $x=\omega-\Delta$ , $\gamma_1^+(\tau)$ reads
\begin{equation}
\begin{aligned}
\gamma_1^+(\tau)=&\frac{2}{\tau}\int_{2\pi/\tau}^\infty dx \frac{G(x+\Delta)}{x^2}-\frac{2 G(\Delta)}{\tau}\int_{2\pi/\tau}^\infty dx \frac{1}{x^2}\\
=&\frac{2A\omega_c^{1-s}}{\tau}\int_{2\pi/\tau}^\infty dx \frac{(x+\Delta)^s e^{-(x+\Delta)/\omega_c}}{x^2}-\frac{G(\Delta)}{\pi}\\
\approx &\frac{2A\omega_c^{1-s}}{\tau}\int_{2\pi/\tau}^\infty x^{s-2} e^{-x/\omega_c}dx-\frac{G(\Delta)}{\pi}\\
=&\frac{2A}{\tau}\Gamma\left(s-1,\frac{2\pi}{\omega_c\tau}\right)-\frac{G(\Delta)}{\pi}\\
\approx &\frac{2A\Gamma(s-1)}{\tau}-\frac{G(\Delta)}{\pi}+{\it O}(\tau^{-s}).
\end{aligned}
\end{equation}
where the third line is based on the approximation $\Delta/\omega_c\approx 0$ and the last line takes advantage of the series representation of the incomplete $\Gamma$ function $\Gamma(\alpha,x)=\int_x^\infty t^{\alpha-1}e^{-t}dt$~\cite{Daniel2015},
\begin{equation}
\Gamma(\alpha,x)=\Gamma(\alpha)-\sum_{n=0}^\infty \frac{(-1)^n x^{\alpha+n}}{n!(\alpha+n)}\quad[\alpha\neq 0,-1,-2,\cdots]
\end{equation}
Similarly,
\begin{equation}
\begin{aligned}
\gamma_1^-(\tau)>&-\frac{2G(\Delta)}{\tau}\int_{-\Delta}^{-2\pi/\tau}\frac{dx}{x^2}
=G(\Delta)\left(\frac{2}{\Delta\tau}-\frac{1}{\pi}\right)\\
>&-\frac{G(\Delta)}{\pi}.
\end{aligned}
\end{equation}
Note $\gamma_1^-(\tau)$ is negative in this case, so that it stays in the scope of $(-G(\Delta)/\pi,0]$. Since $G(\Delta)/\pi/\gamma_0\approx 0.05$ is a small quantity, we finally obtain
\begin{equation}\label{larges}
\frac{\gamma_{\rm eff}(\tau)}{\gamma_0}=1+\frac{\gamma_{1}(\tau)}{\gamma_0}
\approx 1+\frac{A\Gamma(s-1)}{\pi G(\Delta)\tau}+{\it O}(\tau^{-s}).
\end{equation}

When $s>1$, $\gamma_1^+(\tau)$ becomes non-negligible, and when $s\geq 2$, $\gamma_1^+(\tau)$ contributes significantly more than $\ti{\gamma}_1(\tau)$ to $\gamma_1(\tau)$. Yet we still have $\gamma_{1}(\tau)/\gamma_0>0$ because $\Gamma(s-1)>0$ as long as $s>1$. Thus the rigorous result coincides with our result in Sec.~\ref{powerlaw}: the QAZE always dominates for the super-Ohmic environments.

\begin{figure}[htbp]
\centering
\includegraphics[width=0.5\textwidth]{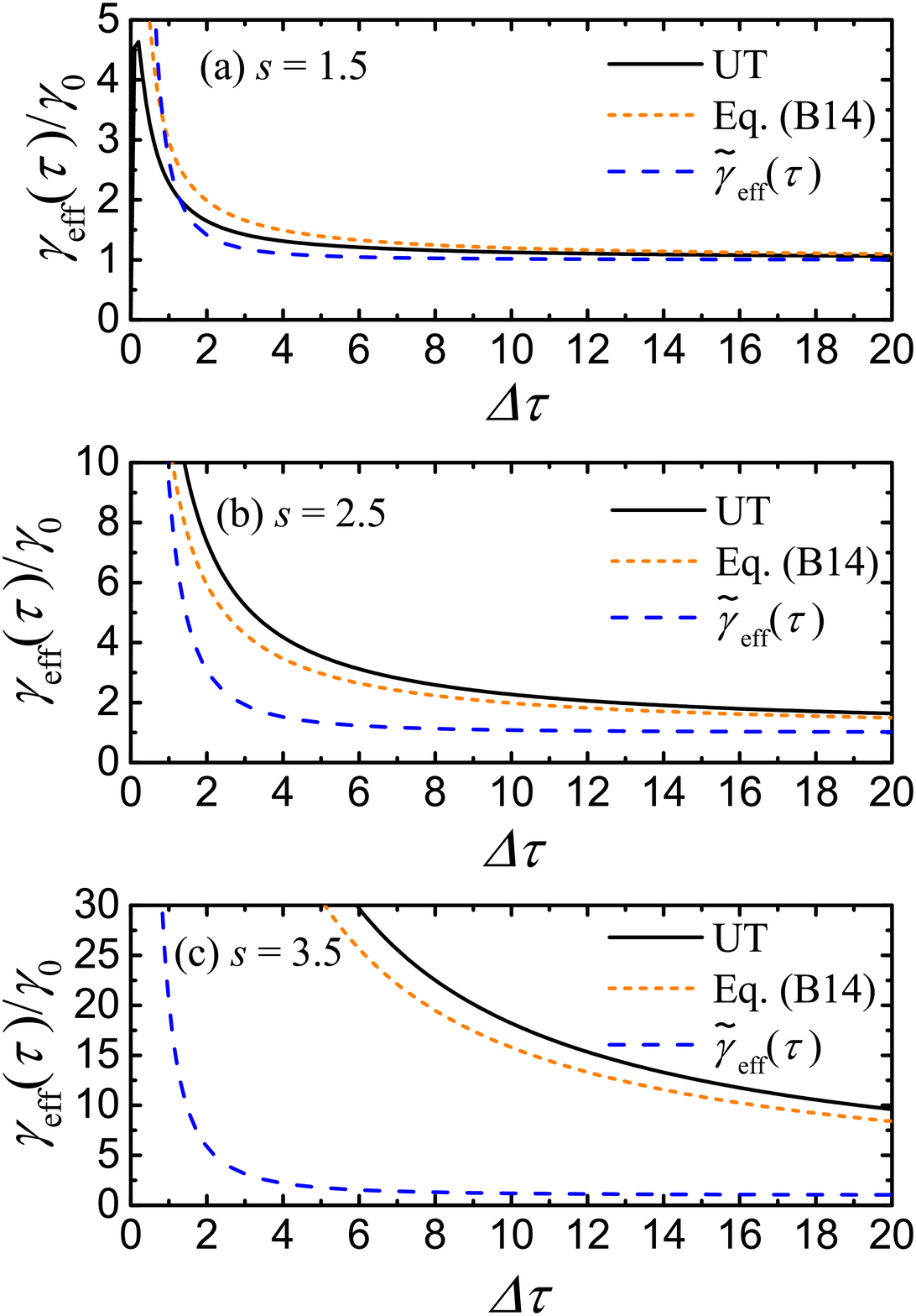}
\caption{(Color online) The effective decay rate vs $\Delta\tau$ for the power-law spectral density function with (a) $s=1.5$, (b) $s=2.5$, and (c) $s=3.5$. The cut-off frequency is chosen as $\omega_c=10\Delta$.}
\label{Appe}
\end{figure}
In physics, the peak of the power-law spectral density function is located at $\omega=s\omega_c$. The distance between $\Delta$ and the center of gravity of $G(\omega)$ increases with the power index $s$ of the SDF, so that the contribution from the minor lobes of the filter function $F(\omega, \tau)$ also increased with $s$. In Fig.~\ref{Appe}, we compare the results by our criterion and those by Eq.~(\ref{larges}) for three super-Ohmic baths with $s>1$. For $s=1.5$, the peak of the spectral function is not extremely larger than $\Delta$. The effective decay rates $\tilde{\gamma}_{\rm eff}(\tau)$ in both Eq.~(\ref{geff}) and Eq.~(\ref{larges}) are closely around those by the UT approach given by Eq.~(\ref{gamma}). For $s=2.5$ and $s=3.5$, the results by Eq.~(\ref{geff}) deviate from those by the UT approach in quantity, and the results by Eq.~(\ref{larges}) (see the orange dotted lines) fit well with them. In these two cases, the contribution from the minor lobes $\gamma_1^+(\tau)$ overwhelms the leading order correction $\tilde{\gamma}_1(\tau)$. In quality, all the results in Fig.~\ref{Appe} still imply the QAZE under super-Ohmic environments, which can be predicted by our criterion.

\section{Decay rate for Lorentzian spectral density}\label{Lorent}

Performing the Laplace transformation on the excited-state amplitude (\ref{alpha}), we get
\begin{equation}
\tilde{\alpha}(p)=\frac{i}{p-\sum_k\frac{|g_k|^2}{p+\Delta-\omega_k}}.
\end{equation}
Consider a finite-band spectrum by taking the SDF $G_L(\omega)$ in the Lorentzian form
\begin{equation}
G_L(\omega)=\sum_k|g_k|^2\delta(\omega-\omega_k)\to\frac{D_0\Lambda^2}{(\omega-\omega_0)^2+\Lambda^2},
\end{equation}
with $\omega_0$ the spectral center, $D_0$ the spectral height, and $\Lambda$ the spectral halfwidth. Then we have
\begin{equation}
\sum_k\frac{|g_k|^2}{p+\Delta-\omega_k}=\int_{-\infty}^\infty d\omega\frac{G_L(\omega)}{p+\Delta-\omega}.
\end{equation}
Here we have expanded the integration domain from $-\infty$ to $\infty$, which neglects the threshold effects~\cite{Piraux1990}. Thus we have
\begin{equation}
\tilde{\alpha}(p)=\frac{i}{p-\frac{\pi D_0\Lambda}{p+\Delta-\omega_0+i\Lambda}}.
\end{equation}
Using the residue theorem, the time-dependent amplitude can be obtained via the inverse Laplace transform,
\begin{equation}
\alpha(t)=\frac{a_+e^{-ia_-t}-a_-e^{-ia_+t}}{a_+-a_-},
\end{equation}
where $a_\pm=[\Omega-i\Lambda\pm\sqrt{(\Omega-i\Lambda)^2+4\pi D_0\Lambda}]/2$ with $\Omega=\omega_0-\Delta$. Now we have the exact decay rate modified by frequent measurements from Eq.~(\ref{s}):
\begin{equation}
\gamma(\tau)=-\frac{1}{\tau}\ln |a(\tau)|^2=-\frac{2}{\tau}\ln\left|\frac{a_+e^{-ia_-\tau}-a_-e^{-ia_+\tau}}{a_+-a_-}\right|.
\end{equation}

In the short-time limit valid for the QZE, the effective decay rate for Lorentzian spectrum from Eq.~(\ref{reff}) yields
\begin{equation}
\begin{aligned}
\gamma_{\rm eff}(\tau)=&\frac{2D_0\Lambda^2}{\tau}\int_{-\infty}^\infty d\omega \frac{1-\cos[(\omega-\Delta)\tau]}{[(\omega-\omega_0)^2+\Lambda^2](\omega-\Delta)^2}\\
=&\gamma_0 \left[1+\frac{\sin\theta-\sin(\theta+\Omega\tau)e^{-\Lambda\tau}}{\Lambda\tau}\right],
\end{aligned}
\end{equation}
under the UT approach. Here $\theta$ satisfies $\sin\theta=(\Omega^2-\Lambda^2)/(\Omega^2+\Lambda^2)$ and $\cos\theta=2\Omega\Lambda/(\Omega^2+\Lambda^2)$, and $\gamma_0=2\pi G_L(\Delta)=2\pi D_0\Lambda^2/(\Omega^2+\Lambda^2)$.

\section{Effective Hamiltonian without the rotating-wave approximation}\label{WRWA}

The second-order effective Hamiltonian with no rotating-wave approximation~(\ref{nRWA}) can be derived by using the following unitary transformation:
\begin{equation}
H_{\rm eff}=\exp(S)\mathcal{H}\exp(-S)=H_0+H_1+H_2\cdots,
\end{equation}
where
\begin{equation}
S=\sum_k(\xi_k^* b_k^\dag\sigma^+-\xi_k b_k\sigma^-),
\end{equation}
with undetermined coefficients $\xi_k$. In order to eliminate the counter-rotating terms $b_k^\dag\sigma^++\rm H.c.$ from the first-order term, we require
\begin{equation}
\begin{aligned}
H_1=&\mathcal{H}_{I}+[S,H_0]\\
=&\sum_k\left[g_k^*(b_k^\dag+b_k)\sigma^+-(\Delta+\omega_k)\xi_k^*b_k^\dag\sigma^++\rm H.c.\right]\\
=&\sum_k\left(g_k^*b_k\sigma^++g_kb_k^\dag\sigma^-\right).
\end{aligned}
\end{equation}
The preceding equation gives rise to the coefficients $\xi_k=g_k/(\Delta+\omega_k)$ and then yields the anti-Hermitian operator $S$ in Eq.~(\ref{HS}). If we neglect the high-frequency intercrossing terms such as $b_k^\dag b_{k'}^\dag$ and $b_kb_{k'}$, we can obtain
\begin{equation}
\begin{aligned}
H_2=&\left[S,\mathcal{H}_{I}\right]+\frac{1}{2}[S,[S,H_0]]\\
=&\sum_k\frac{|g_k|^2}{\Delta+\omega_k}\sigma^+\sigma^--\sum_k\frac{|g_k|^2}{\Delta+\omega_k}.
\end{aligned}
\end{equation}
Thus, the effective Hamiltonian can be written as
\begin{equation}
H_{\rm eff}=\Delta_1\sigma^+\sigma^-+\sum_k\omega_kb_k^\dag b_k
+\sum_k(g_k^*b_k\sigma^++\rm H.c.),
\end{equation}
where $\Delta_1=\Delta+\sum_k|g_k|^2/(\Delta+\omega_k)$. The constant term has been omitted, which has no effect on the dynamical evolution.

\bibliographystyle{apsrevlong}
\bibliography{QZE}

\end{document}